\documentclass[prb,twocolumn,floatfix,showpacs,showkeys,a4paper]{revtex4}
\usepackage{graphicx}
\usepackage[TABTOPCAP]{subfigure}
\usepackage{bm}
\usepackage{amsfonts}
\usepackage{amsmath}
\usepackage{amssymb}
\usepackage{dcolumn}

\begin{document}

\title{Coexistence of magnetism and superconductivity in a \mbox{$t$-$J$} 
bilayer}

\author{Amal Medhi}
\email{amedhi@iitg.ernet.in}
\author{Saurabh Basu} %
\email{saurabh@iitg.ernet.in}%
\author{C.Y. Kadolkar}
\email{charu@iitg.ernet.in}%
\affiliation{Department of Physics, Indian Institute of Technology
Guwahati, Guwahati, Assam 781039, India } %
\date{\today}

\begin{abstract}
We investigate coexistence of antiferromagnetic and superconducting correlations
in bilayered materials using a two-dimensional \mbox{$t$-$J$} model with
couplings across the layers using variational Monte Carlo calculations. It is
found that the underdoped regime supports a coexisting phase, beyond which
the ($d$-wave) superconducting state becomes stable. Further, the effects
of interplanar coupling parameters on the magnetic and superconducting
correlations as a function of hole doping are studied in details. The magnetic
correlations are found to diminish with increasing interplanar hopping  
away from half filling, while the exchange across the layers strengthens
interplanar antiferromagnetic correlations both at and away from half filling.
The superconducting correlations show more interesting features where larger 
interplanar hopping considerably reduces planar correlations at
optimal doping, while an opposite behaviour, {\it {i.e.}} stabilisation
of the superconducting state is realised in the overdoped regime, with the 
interplanar exchange all the while playing a dormant role.
\end{abstract}


\maketitle

\section{Introduction}
The mechanism leading to electron pairing in copper oxide superconductors
has stimulated a great deal of speculation. The manifestation of planar 
antiferromagnetism in the ${\rm CuO_{2}}$ layers have provided motivation
to study the role and importance of antiferromagnetic (AF) interactions between
${\rm {{Cu}^{2+}}}$ spins and their intimate relevance to superconductivity
\cite{AndersonSCI87}. Doping with holes in these insulating cuprates
results in destruction of long range order, however short range 
antiferromagnetic correlations between the copper moments survive. Further
doping leads to the emergence of a superconducting (SC) state.

Thus the interplay of AF and SC phases has generated much attention. A 
phenomenological $SO(5)$ theory has attempted to unify antiferromagnetism and 
superconductivity owing to their 
proximity in the phase diagram\cite{ZhangSCI97, DemlerRMP04}. 
The basic assumption of the theory is that these two phases 
share a common microscopic origin and hence both demand treatment at an equal 
footing. A number of experimentally observed features, such as the vortex state,
a resonant peak in SC state from neutron scattering data for optimally and 
underdoped samples etc.\ have provided ample credence to the theory. However a 
microscopic theory in this regard is still lacking.

A natural extrapolation of finding the connection between AF and SC is the
issue of their coexistence which seems more crucial and calls for attention.
A large volume of work exists that focuses on the various details of the
coexistence phenomenon. Some of the theoretical attempts include mean field
studies of \mbox{$t$-$J$} and Hubbard-like models which confirm the coexistence of 
magnetic and superconducting order \cite{InuiPRB88,YamasePRB06}. The coexistence is suggestive
of the presence of short range AF correlations in SC state that are probed by
inelastic neutron scattering experiments via an enhanced scattering
intensity near the AF wavevector $(\pi,\pi)$
\cite{MignodPHC91,DaiPRL98,KotegawaPRB04,BirgeneauPRB88}.

The coexistence issue has been revived recently in the context of bilayer (and 
multilayer) cuprates \cite{YamasePRB06,KotegawaPRB04}. While acknowledging the planar correlations 
perhaps dominate the physical properties of these superconductors, the role of
interlayer couplings and their relevance to the coexistence phenomena for 
layered materials have been heavily emphasized. Intimately connected to this
is the question: whether the superconducting correlations originate from
AF spin fluctuations \cite{MonthouxPRB94,MonthouxPRL94,PaoPRL94} or via a 
electron-phonon mediated pairing enhanced by interlayer 
tunneling\cite{ChakravartySCI93}. However both these mechanisms cannot be operative together. 

The next fundamental question is the symmetry of the superconducting gap 
function. Even with sufficient experimental evidence for a $d_{x^{2}-y^{2}}$
pairing for a planar materials (for a comprehensive review on the subject see
Ref.~\onlinecite{ScalapinoPHYSREP95}), for bilayers the 
pairing symmetry is still unclear. Several variational calculations
\cite{YokoyamaJPSJ88,YokoyamaJPSJ96,HimedaPRB99} performed over the years have provided nourishment
to a $d_{x^{2}-y^{2}}$-wave pairing scenario and a reasonably broad window of
carrier concentration has been identified in planar systems where magnetic 
and SC order coexist. Similar studies in the context
of bilayers are lacking and hence provide motivation for us to investigate a 
bilayer \mbox{$t$-$J$} model via variational Monte Carlo (VMC) technique.

Our goal in this work is to examine the coexistence of antiferromagnetism and
superconductivity in \mbox{$t$-$J$} bilayers using VMC and to study the magnetic
and superconducting correlation in the variational ground state. 
We further intend to investigate the dependence of these properties on interlayer 
coupling strengths. In a recent paper\cite{MedhiPHC06}, to determine the most 
suitable pairing symmetry of the SC state in \mbox{$t$-$J$} bilayers we investigated 
the stability of various pairing symmetries, e.g.
(a) $\Delta(\cos k_{x} - \cos k_{y})$ ($d$-wave),
(b) $\Delta_{||}(\cos k_{x} - \cos k_{y}) + \Delta_{\perp}\cos k_{z}$ and
(c) $\Delta_{||}(\cos k_{x} - \cos k_{y}) + \Delta_{\perp}(1- \cos k_{z})$.
Another function discussed in connection with bilayer materials but not
included in our previous study is the 
$s^{\pm}$ state ($\Delta_{k} = \pm \Delta(\cos k_{x}+\cos k_{y})$
with `$+$' sign for $k_{z} = 0$ and `$-$' for $k_{z} = \pi$) 
which has $s$-wave symmetry and opposite signs in the 
bonding ($k_{z} = 0$) and antibonding bands ($k_{z} = \pi$)
\cite{MazinPRL95,LiechtensteinPRL95}. This state possesses the
merit of explaining the resonance peak at 41 $meV$ obtained in neutron 
scattering experiments \cite{MignodPHC91,MookPRL93,FongPRL95,RegnaultPHC94}.
In this paper we consider also this pairing symmetry in the search for
most stable ground state in bilayers.

Here we state the main results obtained by us. The long range AF order coexists with
superconductivity in the underdoped regime. In the coexisting phase not
only the AF but SC correlations are also significantly stronger as compared to that in the
pure SC ($d$-wave) state. Larger interlayer hopping frequency reduces planar SC correlations
in the optimally doped phase whereas it enhances it in the overdoped regime,
while the effect of interlayer exchange on SC correlations is minimal for the range of
the parameter values considered in our paper.

Our paper is organised as follows: section II introduces the \mbox{$t$-$J$} 
model for
bilayers and discusses the most suitable variational wavefunction to be used
for our calculations. A brief note on the numerics used and an elaborate 
discussion on the results appear in section III. The effects of interplanar
coupling parameters have been emphasized in magnetic, superconducting and
the coexisting phase. Section IV concludes with a brief summary of the results
obtained in this paper.

\section{The Hamiltonian and the variational wavefunction}

The \mbox{$t$-$J$} Hamiltonian for a bilayer can be written as
{\small
\begin{eqnarray}
\label{eq:tj_model}
\mathcal{H} = - t\sum_{\langle i,j\rangle\sigma}\left(c_{i\sigma}^\dag 
c_{j\sigma} + H.c.\right)+J\sum_{\langle i,j\rangle}\left( {\bf S}_i.{\bf S}_j 
- \frac{1}{4}n_i n_j \right) \nonumber \\
-t_{\perp}\sum_{\langle\langle i,k\rangle\rangle\sigma}\left(c_{i\sigma}^\dag 
c_{k\sigma} + H.c.\right) 
+J_{\perp}\sum_{\langle\langle i,k\rangle\rangle}\left({\bf S}_i.{\bf S}_k-
\frac{1}{4}n_i n_k \right)
\end{eqnarray}}
where $t$ and $J$ are the planar hopping and exchange integral respectively,
while $t_{\perp}$ and $J_{\perp}$ are the corresponding interplanar 
parameters. $c_{i\sigma}$ ($c_{i\sigma}^\dag$) annihilates (creates) an 
electron of spin $\sigma$ at site $i$, $n_{i} = 
\sum_{\sigma}c_{i\sigma}^\dag c_{i\sigma}$ and $\mathbf{S}_i$ is the spin
operator at site $i$ given by 
$S_i^\alpha = \psi_i^\dag (\frac{1}{2}\bm{\sigma}_{\alpha})\psi_i$.
$\psi_i^\dag = \left(c_{i\uparrow}^\dag\;\; c_{i\downarrow}^\dag\right)$
and $\bm{\sigma}_{\alpha}$ (with $\alpha = x, y, z$) are the Pauli spin 
matrices. The summation indices $\langle i,j\rangle$ and $\langle\langle i,k
\rangle \rangle$ indicate nearest neighbour pairs in the same plane and 
different planes respectively. The Hamiltonian obeys an essential requirement, 
{\it {i.e.}} it acts on a subspace of no doubly occupied sites.

To incorporate the coexistence of AF and SC phases we consider the following 
variational wavefunction as the ground state of the Hamiltonian. 
\begin{eqnarray}
|\Psi_{var}\left(\Delta_{sc},\Delta_{af}\right)\rangle = \mathcal{P}_G\mathcal{P}_N
\prod_\mathbf{k}\left(u_\mathbf{k}+v_\mathbf{k}d_{\mathbf{k}\uparrow}^{\dag}
d_{-\mathbf{k}\downarrow}^{\dag}\right)|0\rangle
\label{eq:variational_wf}
\end{eqnarray}
where the operator $\mathcal{P}_N$ projects out the states with 
a fixed electron number, $N$ and $\mathcal{P}_G=\prod_i(1-n_{i\uparrow}n_{i\downarrow})$, 
is the Gutzwiller projector which imposes the condition of no double occupancy.  
The product in Eq.~\ref{eq:variational_wf} is for over all the `$\mathbf{k}$' 
points in  the first Brillouin zone and the amplitudes $u_{\mathbf{k}}$ 
and $v_{\mathbf{k}}$ are defined by,
\begin{eqnarray}
\frac{v_\mathbf{k}}{u_\mathbf{k}} = 
\phi(\mathbf{k})=\frac{\Delta_\mathbf{k}}{\left(\mp E_{\mathbf{k}}-\mu\right) 
+ \sqrt{\left( \mp E_\mathbf{k}-\mu\right)^2+\Delta_\mathbf{k}^2}} 
\label{eq:phik}
\end{eqnarray}
where $\Delta_\mathbf{k}=\Delta_{sc}f(\mathbf{k})$ represents the SC gap, 
$f(\mathbf{k})$ being an appropriate symmetry function of $\mathbf{k}$ 
and $E_\mathbf{k}=\sqrt{\epsilon_\mathbf{k}^2+\Delta_{af}^2}$. 
$\epsilon_\mathbf{k}=-2t(\cos k_x+\cos k_y)-2t_{\perp}\cos k_z$, is the free
electron dispersion and $\mu$ is the chemical potential. The $-(+)$ signs in
the denominator of Eq.~\ref{eq:phik} corresponds to $\epsilon_\mathbf{k}<0$ 
$(\epsilon_\mathbf{k}>0)$.
The quasiparticle operators, $d_{\mathbf{k}\sigma}^{\dag}$ diagonalizes the 
AF Hartree-Fock Hamiltonian\cite{HimedaPRB99} with a gap
$\Delta_{af}$ and are related to the electron operators by the following 
transformation,
\begin{eqnarray}
\label{eq:transformation}
\left[\begin{array}{l} 
d^{\dag}_{\mathbf{k}\sigma} \\ d^{\dag}_{\mathbf{k+Q}\sigma} \end{array}\right]
= \left[\begin{array}{cc} 
\alpha_\mathbf{k} & \eta(\sigma)\beta_\mathbf{k} \\
-\eta(\sigma)\beta_\mathbf{k} & \alpha_{\bf k} 
\end{array}\right]
\left[\begin{array}{l} c^{\dag}_{\mathbf{k}\sigma}\\c^{\dag}_{\mathbf{k+Q}\sigma} 
\end{array}\right]
\end{eqnarray}
with 
\begin{eqnarray}
\label{eq:alpha_beta}
\alpha_\mathbf{k}=\frac{1}{\sqrt{2}}{\left(1-\frac{\epsilon_\mathbf{k}}{E_\mathbf{k}}
\right)}^{1/2} \quad
\beta_\mathbf{k}=\frac{1}{\sqrt{2}}{\left(1+\frac{\epsilon_\mathbf{k}}{E_\mathbf{k}}
\right)}^{1/2}
\end{eqnarray}
Here ${\bf Q} = (\pi, \pi, \pi)$, is the perfect nesting vector and 
$\eta(\sigma) = \pm 1$ for $\sigma = \uparrow,\downarrow$.  

The wavefunction in Eq.~\ref{eq:variational_wf} consists of two 
variational parameters, viz.\ $\Delta_{sc}$ and $\Delta_{af}$.  Ideally the 
chemical potential, $\mu$ should also be treated as a variational parameter, 
however here we fix it at its noninteracting value, $\mu_0$.  This is because 
the energy correction obtained by varying $\mu$ has been found to be negligibly 
small (for small $J$)\cite{YokoyamaJPSJ96} for a square lattice and we expect 
it to be the same also for bilayers. The wavefunction describes different 
phases depending upon the (relative)
values of the variational parameters. For example, $\Delta_{af}=0$ describes 
the usual BCS superconducting state\cite{ParamekantiPRB04}, whereas in 
the limit $\Delta_{sc}\rightarrow 0$, the wavefunction reduces to a state with 
antiferromagnetic long range order\cite{YokoyamaJPSJ87-1}. For nonzero 
$\Delta_{sc}$ and $\Delta_{af}$,
the wavefunction describes a phase with coexisting AF and SC state, 
while the normal state is recovered as both the parameters vanish.

\section{The Results}

We shall skip details of the variational Monte Carlo method used as it appears
elsewhere \cite{YokoyamaJPSJ87-2,ParamekantiPRB04,MedhiPHC06} and only provide 
the essential features of our computation. 
We consider periodic boundary condition along the planar $x$-direction
and antiperiodic boundary condition in the planar $y$-direction to
avoid singularity in $\phi(\mathbf{k})$ for the $\mathbf{k}$-points
with $k_{x} = k_{y}$ and $\epsilon_{\mathbf{k}}-\mu \le 0$ 
\cite{ParamekantiPRB04}.
In one Monte Carlo Sweep (MCS) through the lattice, $N_s$ (equal to the number 
of lattice sites) random moves are attempted which consists of moving an 
electron to an empty site and exchanging two antiparallel spins. 
After each successful move, Monte Carlo updates of the configurations 
are made by using the inverse update method\cite{CeperleyPRB77}. Various
expectation values that are obtained in the paper are computed by sampling 
configurations chosen from about $10^4$ - $10^5$ MCSs after taking 5,000 warm 
up sweeps.  Simulations are performed on a bilayered lattice of size 
\mbox{$10\times 10\times 2$}. 

In the following, we discuss the stability of different phases with different
SC pairing symmetries, followed by the results for magnetic and 
superconducting correlations. We show that the AF and SC phases coexist in the
underdoped regime by comparing the optimal energy of the  variational 
wavefunction having pure SC correlations with the one having coexisting AF 
and SC order. To characterize the pure SC and the coexisting phase and to
enumerate the differences in their properties, we compute the correlation 
functions for both these phases and make a detailed comparison between the two.

\subsection{Stability of different phases}

We first consider the pure SC wavefunction 
(obtained by putting $\Delta_{af}=0$ in Eq.~\ref{eq:variational_wf}). 
As for the pairing symmetry of the SC state, we considered energies of four 
different variational wavefunctions listed in the previous section.  An earlier
work \cite{MedhiPHC06} investigates in details the first three wavefunctions 
in the list and found that the pure $d$-wave state yields lowest energy at all 
values of $\delta$ away from half filling. This
led to the conclusion that of the three pairing symmetries discussed there,
a planar $d$-wave state is most appropriate in the context
of a \mbox{$t$-$J$} bilayer. We have included the $s^{\pm}$-state for
comparison with the existing ones in the light of the emphasis given to
it where it is claimed to be more stable than that of the $d$-wave state
for the bilayered systems and is capable of explaining the origin of the
neutron scattering peak observed experimentally in 
$\rm {YBa_{2}Cu_{3}O_{7}}$ at temperatures below $T_{c}$ 
\cite{MazinPRL95,LiechtensteinPRL95,RadkePRB96}. However our VMC calculations 
indicate  that the energy of the $s^{\pm}$ pairing state to be actually higher 
than the normal state at all values of hole concentrations. Thus we discard 
this pairing symmetry from the list of possible candidates and conclude that
a planar $d$-wave state is most appropriate to study superconductivity in 
bilayers. 

\subsection{Ground state Energy}

Before we proceed to calculate physical quantities, such as the ground state
energy etc., it is somewhat interesting to look at the variation of the 
optimal superconducting variational parameter, $\widetilde{\Delta}_{sc}$, as a
function of hole concentration for a few representative values of the
interplanar coupling parameters. The choices of these parameters are chosen
from experimental data \cite{Rossat-MignodPHB91,ShamotoPRB93}.
\begin{figure}[htb]
\centering
\subfigure[ ]{\label{sf1:optparam_delta}
\includegraphics[width=1.6in,height=1.4in]{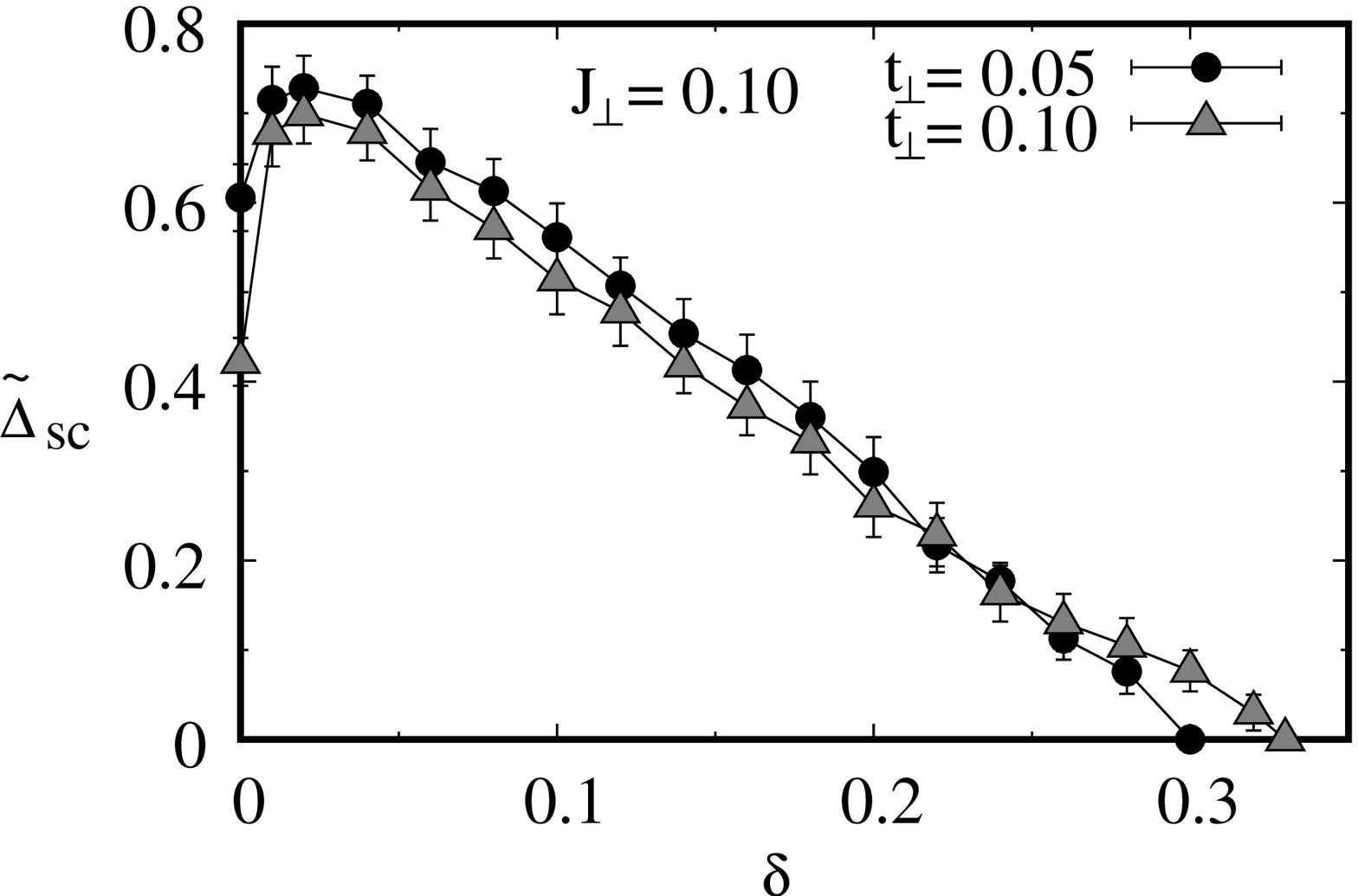}}
\hfill
\subfigure[ ]{\label{sf2:optparam_delta}
\includegraphics[width=1.6in,height=1.4in]{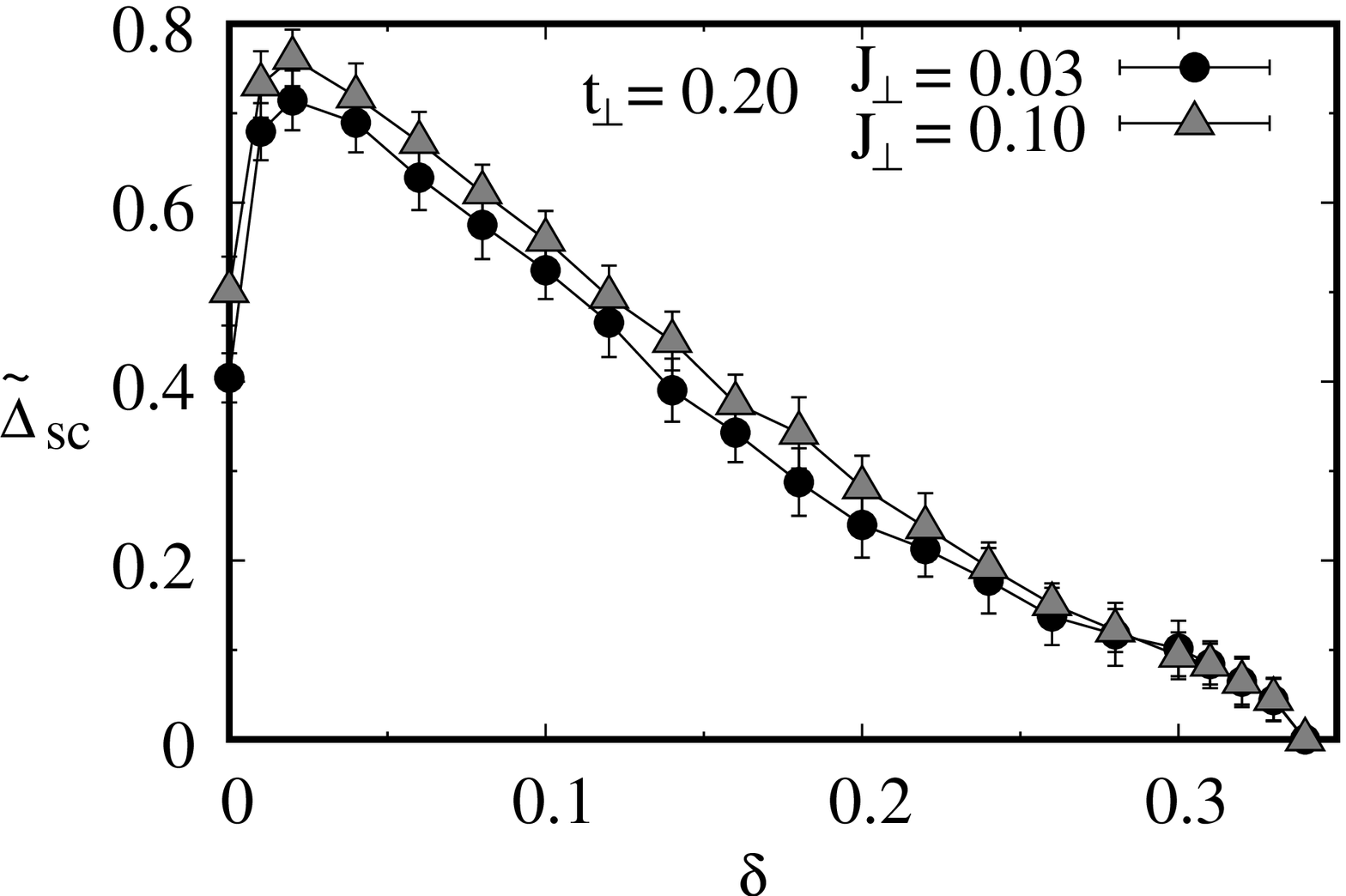}}
\caption{Optimal value of the variational parameter, $\widetilde{\Delta}_{sc}$ 
shown as a function of hole
concentration, $\delta$ corresponding to different choices for the interplanar
parameters. All the figures in the paper correspond to calculations done
on a lattice size of $10\times 10\times2$. The parameter values are taken in
units of (planar) hopping $t$.}
\label{fig:optparam_delta}
\end{figure}
From Fig.~1 it may  be noticed that the critical hole concentration, $\delta_c$
at which $\widetilde{\Delta}_{sc}$ vanishes, increases (from $\sim 0.3$ to 
$\sim 0.34$) with larger interplanar hopping, $t_{\perp}$, while the 
interplanar exchange, $J_{\perp}$ has no significant effect on $\delta_c$. 
Below $\delta_c$, $\widetilde{\Delta}_{sc}$ is slightly reduced by both 
higher values of $t_\perp$ and $J_{\perp}$. It is worth mentioning here that 
in a two dimensional square lattice (with same values for the planar 
parameters), a $d$-wave state is stabilized upto $\sim 28\%$
hole concentration\cite{YokoyamaJPSJ96}, which is lower than the corresponding 
value {\it {i.e.}} 34\% obtained here for the bilayer (Fig.~1). Thus stability 
of the superconducting state extends upto {\em {higher}}  values of hole 
concentration in bilayers than in planar materials.

\begin{figure}[htb]
\centering
\includegraphics[width=3.0in]{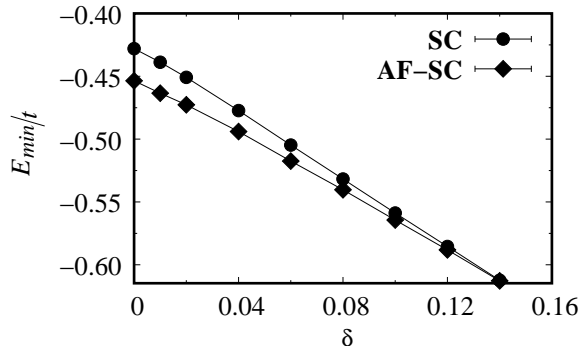}
\caption{Optimal energy, $E_{min}/t$ (per site) for the pure $d$-wave SC state 
and the coexisting AF and SC state as a function of hole concentration, 
$\delta$ for $t_\perp = 0.20$ and $J_{\perp}=0.10$.}
\label{fig:optenergy_wf}
\end{figure}

Next we introduce the second variational parameter, {\it {i.e.}} 
$\Delta_{af}$ into the problem and carry out minimization of  energy in 
two-variational parameters space, $\Delta_{sc}$ and $\Delta_{af}$. 
The calculation shows that the energy is significantly lowered in the
underdoped regime ($\delta<0.14$) when compared to that obtained for the 
pure SC state. The optimal energy, $E_{min}/t$ (per site) as function of 
$\delta$ for the two cases is shown in Fig.~\ref{fig:optenergy_wf} for one 
one particular choice for the interplanar parameters. 
The energy difference between the two phases is maximum at half-filling and
decreases gradually with increasing hole concentration, finally vanishing
at $\delta\sim 0.14$. 
Thus superconductivity coexists with antiferromagnetism in the underdoped 
region for a bilayer, a feature also observed for the two dimensional 
\mbox{$t$-$J$} model\cite{HimedaPRB99}. This is one of the key results
of our paper.  Similar energy difference of the two phases are found 
for other choices of $t_{\perp}$ and $J_{\perp}$ (as appear in Fig.~1). 

\subsection{Magnetic order}
\begin{figure}[htb]
\centering
\subfigure[~Planar spin correlations]{\label{sf1:spincorr_wfSC}
\includegraphics[width=1.6in,height=1.28in]{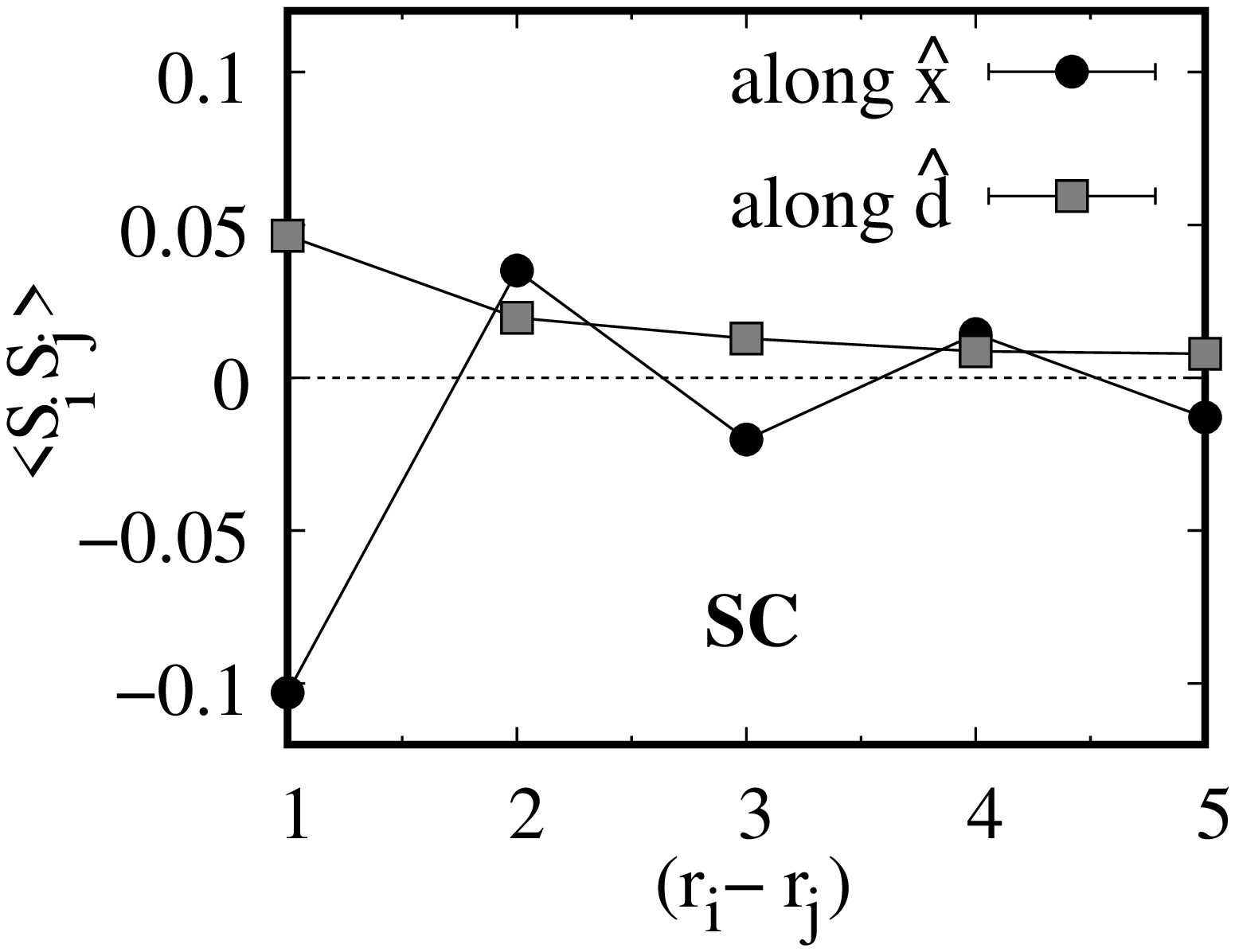}}
\hfill
\subfigure[~Spin structure factor]{\label{sf2:spincorr_wfSC}
\includegraphics[width=1.6in,height=1.28in]{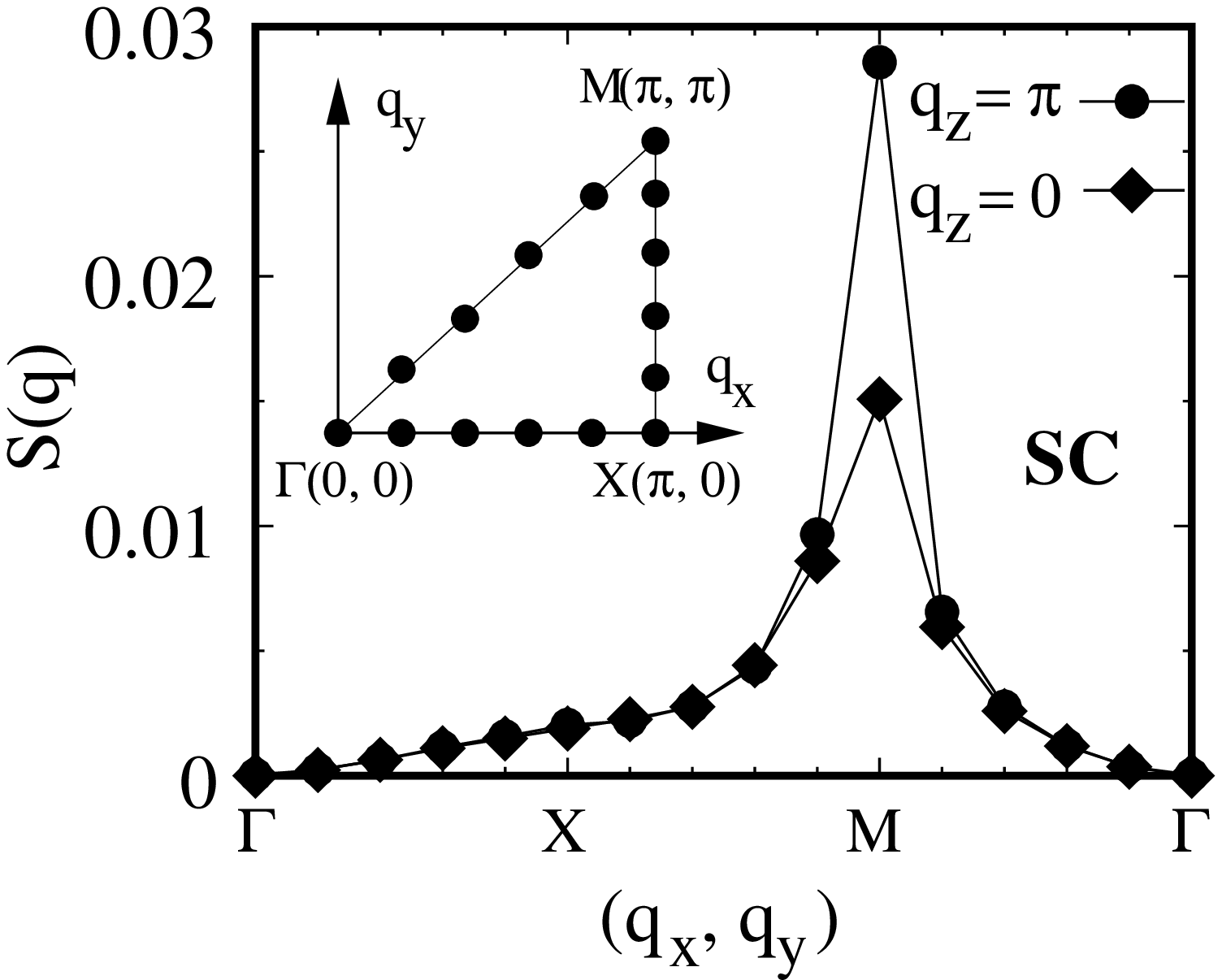}}
\caption{(a) Spin-spin correlation as a function of distance along a planar 
edge ($\hat{x}$-direction) and along a planar diagonal ($\hat{d}$-direction).
(b) Spin structure factor, $S(\mathbf{q})$ as a function of $\mathbf{q}$ chosen
along the symmetry lines shown in figure. The plots shown are at half-filling 
and for $t_\perp = 0.20$ and $J_{\perp}=0.10$.}
\label{fig:spincorr_wfSC}
\end{figure}

We first examine the magnetic correlations in the pure $d$-wave state.
The relevant quantities to compute are planar spin-spin correlations, 
$\langle S_i^zS_j^z\rangle$  and the corresponding Fourier transform,
$S(\mathbf{q})$, called as the structure factor which is defined as,
\begin{eqnarray}
S(\mathbf{q})=\frac{1}{N}
\sum_{ij}e^{i\mathbf{q}.(\mathbf{r}_i-\mathbf{r}_j)}\langle S_i^zS_j^z\rangle
\label{eq:ssf}
\end{eqnarray}
The real-space correlations (Fig.~\ref{sf1:spincorr_wfSC}) shows signature of 
AF order in the planes. 
However the correlations are found to decay as a function of planar distances 
suggesting  an absence of AF long range order (AFLRO) even at half filling. 
It may be noted that the $d$-wave state does not show long range magnetic order
also in two dimensional systems. As for the interplanar correlations, the 
$d$-wave shows AF ordering between the spins in the two layers, however 
the correlations are found to be very weak. For instance, the strength of the
nearest neighbour spin correlation for two corresponding sites in different
planes is approximately 
$10\%$ of that for two sites in the same plane.  
In Fig.~\ref{sf2:spincorr_wfSC}, we plot $S(\mathbf{q})$ as a function of 
$(q_x,q_y)$, the points being chosen along a symmetry path for $q_z$ equal to 
both $0$ and $\pi$. The peak in $S(\mathbf{q})$ at $(\pi,\pi,\pi)$ indicates
the existence of antiferromagnetic correlations in the lattice. However, the 
peak at $(\pi,\pi,0)$ is of comparable magnitude to that at $(\pi,\pi,\pi)$,
which corroborates the presence of weak interplanar correlations mentioned 
earlier \cite{Weakcorrelations}. Away from half-filling, both planar and 
interplanar spin correlations are found to vanish rapidly with hole doping. 
Other values for the interplanar parameters among the ones considered here 
are found to have little impact on spin correlations except for a small reduction 
of planar correlations by larger interplanar hopping away from half filling. 
The value of staggered structure factor, $S(\pi,\pi,\pi)$ at half-filling 
for the $d$-wave state is obtained as $\sim 0.03$. This value may be compared 
with the exact diagonalization results for a bilayer \mbox{$t$-$J$} 
model\cite{EderPRB95}
which yields an estimate for $S(\pi,\pi,\pi)$ to be $\sim 0.45$ for 
$J_{\perp}/J \simeq 0.28$ (same corresponding value is used in 
Fig.~\ref{fig:spincorr_wfSC}) at half filling\cite{CompareSSF}. Thus the 
staggered magnetization for the pure $d$-wave state obtained in our calculations is 
far lower than that obtained via exact diagonalization studies. 
This large discrepancy can be attributed mainly to the absence of interlayer 
magnetic correlations for the pure $d$-wave state. 

\begin{figure}[htb]
\centering
\subfigure[~Planar correlations]{\label{sf1:spincorr_wfAFSC}
\includegraphics[width=1.6in]{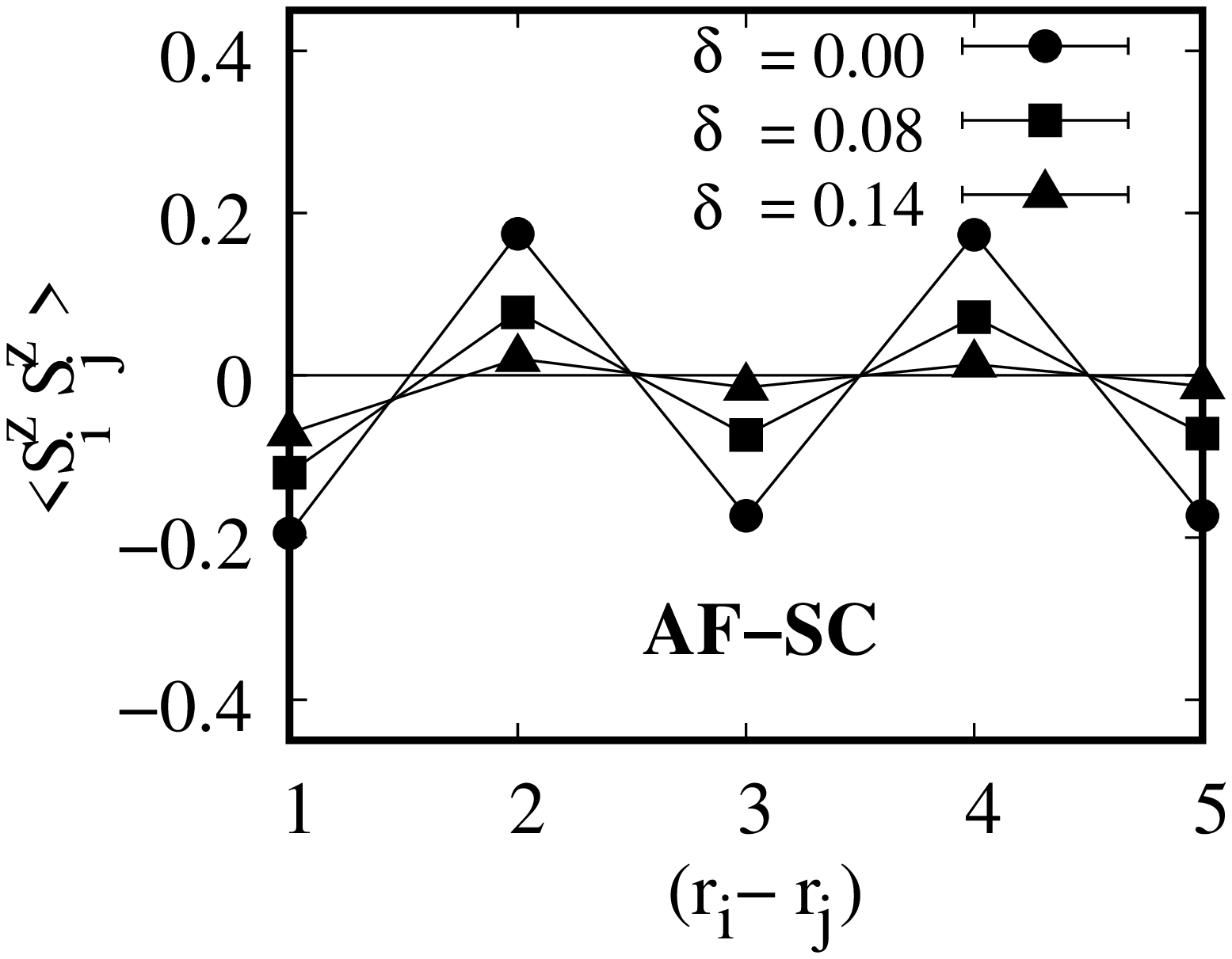}}
\hfill
\subfigure[~Interplanar correlations]{\label{sf2:spincorr_wfAFSC}
\includegraphics[width=1.6in]{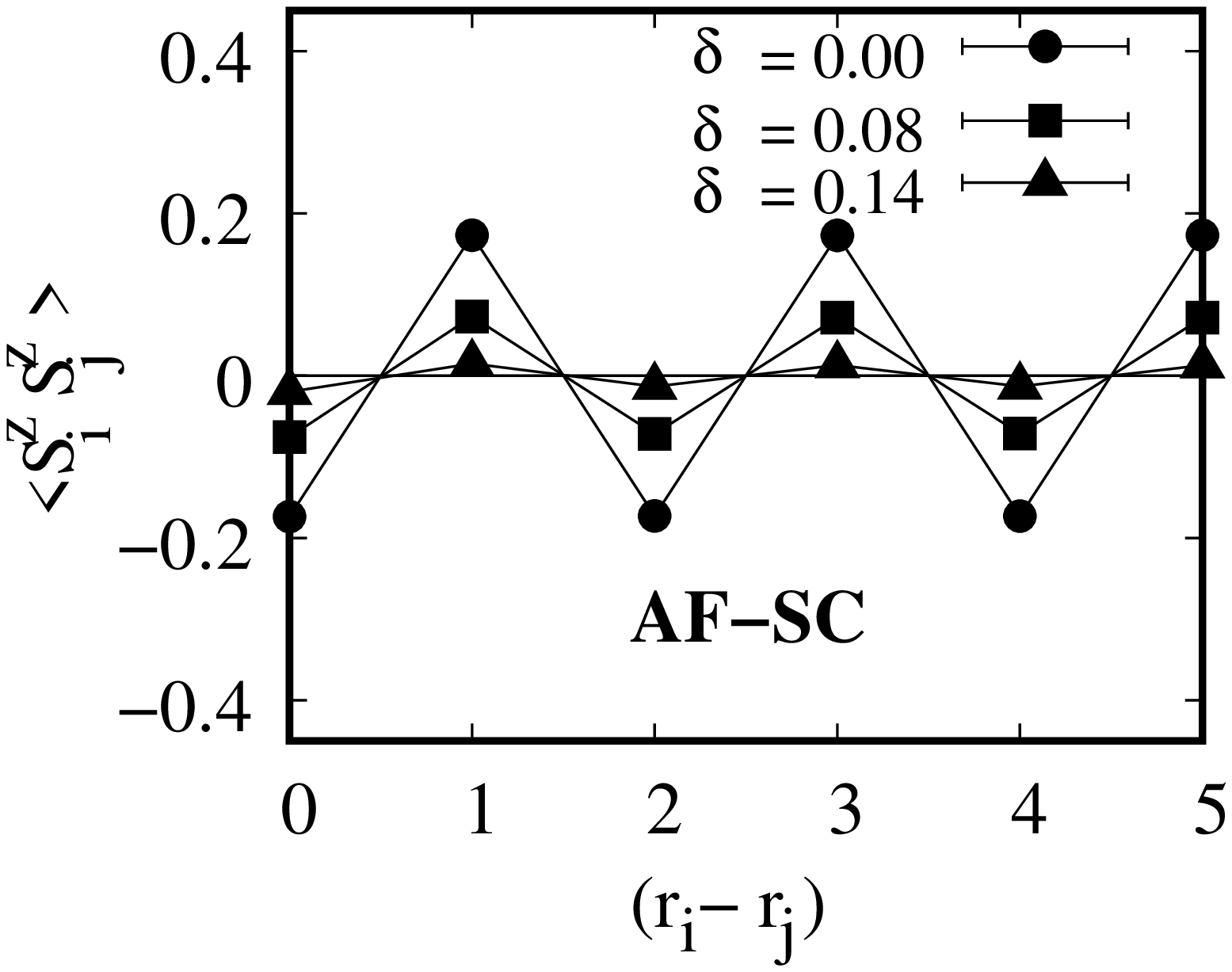}}
\caption{Spin spin correlation as a function of distance along 
$\hat{x}$-direction for the coexisting AF and $d$-wave SC (AF-SC) state. The 
spins are on lattice sites lying (a) along an edge in a plane, and
(b) on two different planes along the same direction. The values of the
hole concentration, $\delta$ are shown in the figures. Here the interplanar
parameters are again chosen as $t_\perp = 0.20$ and $J_{\perp}=0.10$.}
\label{fig:spincorr_wfAFSC}
\end{figure}

\begin{figure}[htb]
\centering
\subfigure[~$q_z = \pi$]{\label{sf1:ssf_wfAFSC}
\includegraphics[width=1.6in]{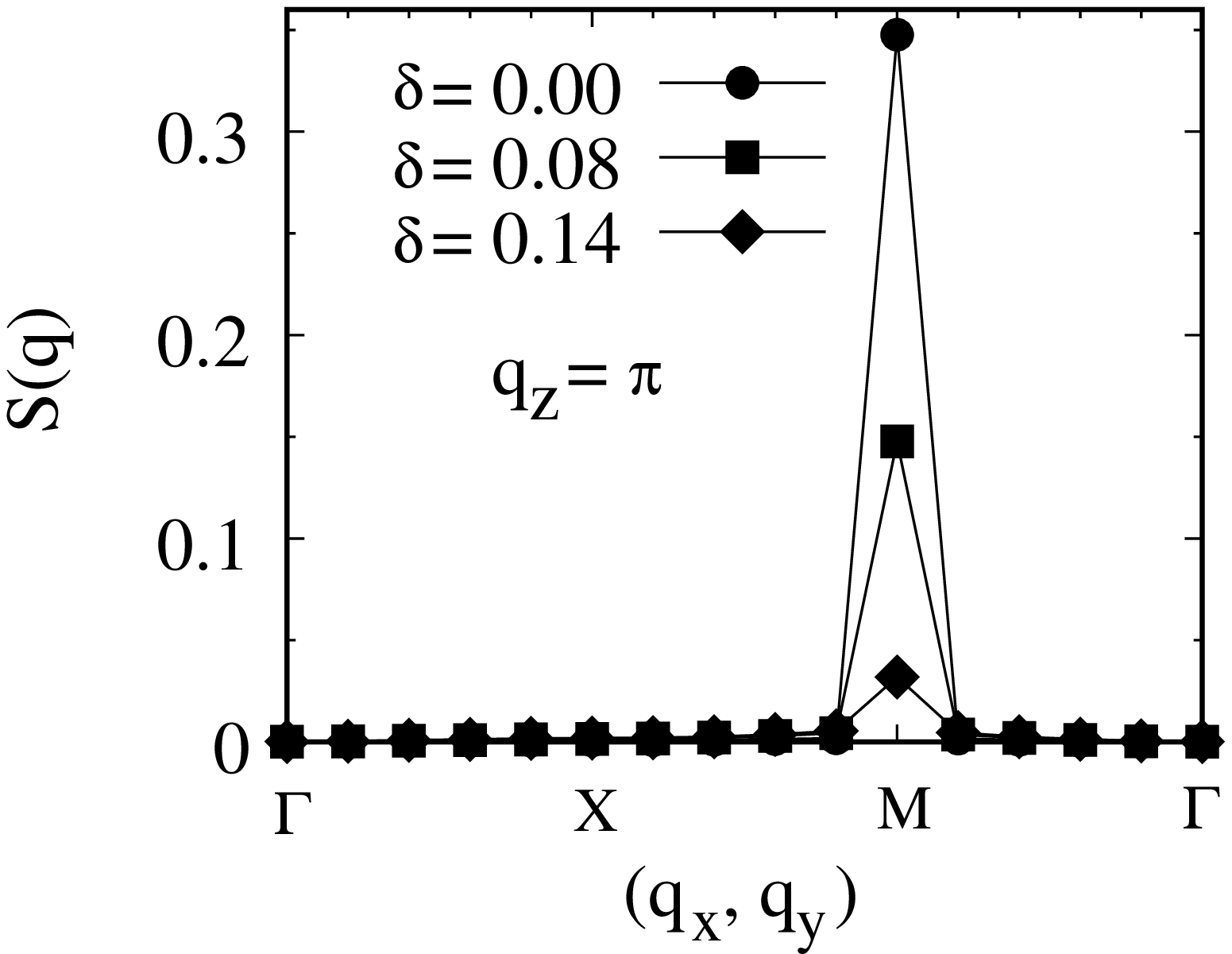}}
\hfill
\subfigure[~$q_z = 0$]{\label{sf2:ssf_wfAFSC}
\includegraphics[width=1.6in]{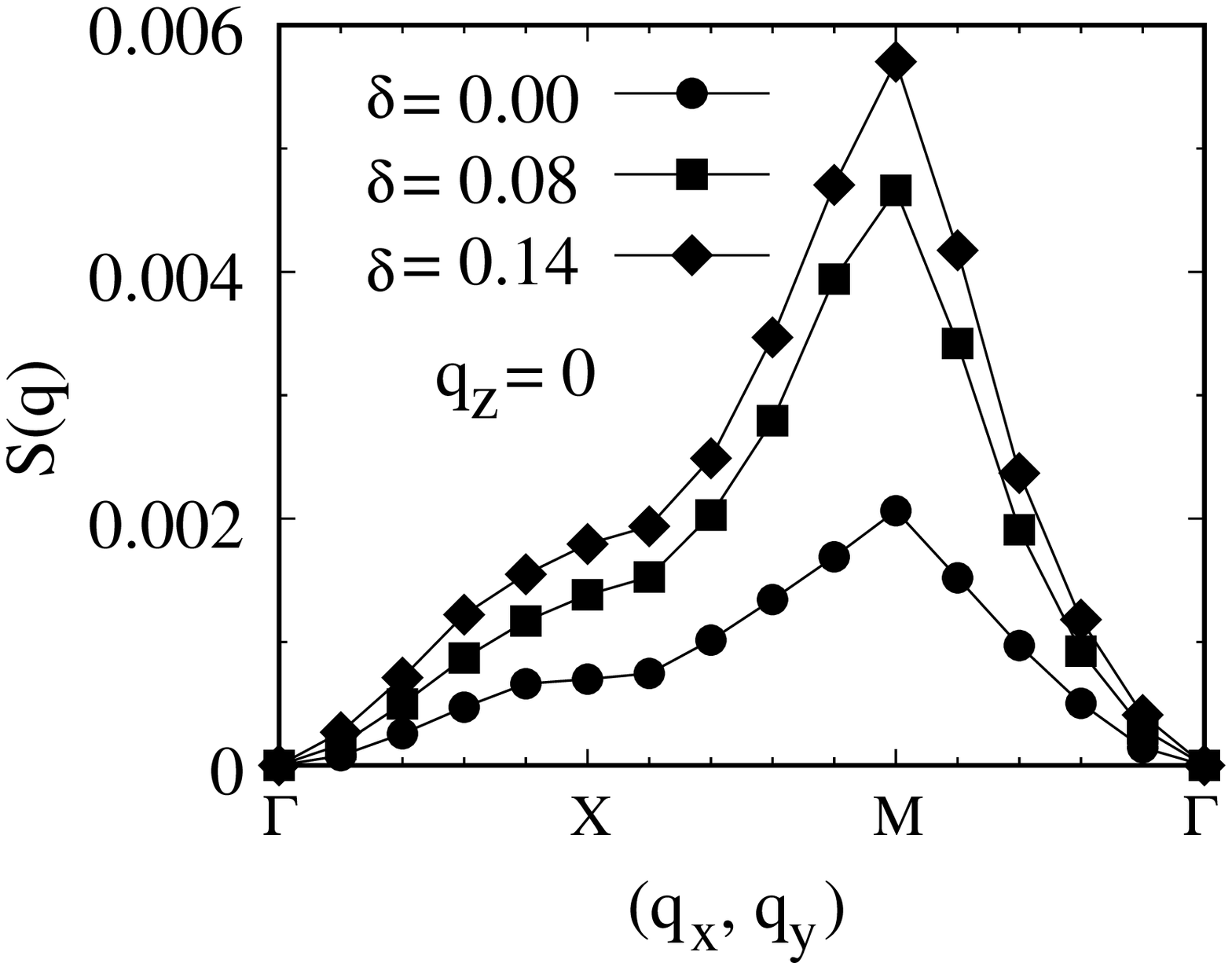}}
\caption{$S(\mathbf{q})$ as a function of $\mathbf{q}$ for the AF-SC 
wavefunction at various hole concentrations shown in figure. 
The $(q_x,q_y)$ points are chosen in the same way as in Fig.~\ref{sf2:spincorr_wfSC}.   
$t_\perp = 0.20$ and $J_{\perp}=0.10$.}
\label{fig:ssf_wfAFSC}
\end{figure}

Next we discuss the magnetic correlations in the coexisting AF and $d$-wave SC 
state which is lower in energy than that of the pure $d$-wave SC state in the 
underdoped region. In Fig.~\ref{fig:spincorr_wfAFSC}, we show the planar and
interplanar spin correlations both at and away from half-filling. 
The correlations are clearly much stronger in this case than for the pure
$d$-wave state. The interplanar correlations, which was very weak in 
the $d$-wave state, is almost of the same magnitude as the planar correlations.  
In addition, the magnitude of correlations does not seem to decay with distance at and
even slightly away from half-filling. However, at larger values of $\delta$,
the magnetic correlations decay rapidly with distance as seen for $\delta=0.14$ 
in Fig.~\ref{fig:spincorr_wfAFSC}. The energy calculations also provide a 
support for this result where it is found that the kinetic energy dominates 
over the exchange energy and consequently the antiferromagnetic phase 
disappears corresponding to $\delta \sim 0.14$. The presence of magnetic order 
in the coexisting phase is further emphasized by plotting $S(\mathbf{q})$
versus $\mathbf{q}$ at various hole concentrations in 
Fig.~\ref{fig:ssf_wfAFSC}. 
The sharp peaks in $S(\mathbf{q})$ at ${\mathbf{q}}=(\pi,\pi,\pi)$ for small values 
of hole doping (Fig.~\ref{sf1:ssf_wfAFSC}) indicates the existence strong AF long 
range correlations in the system. It should be noted that the value of 
$S(\pi,\pi,\pi)$ at half-filling is $\sim 0.35$ which is considerably 
larger than that obtained for the pure SC state and is comparable to the exact 
diagonalization value ({\it {viz.}} $0.45$) mentioned 
earlier \cite{EderPRB95}. Thus the support for the coexisting phase becomes
more robust. Also $S(\pi,\pi,\pi)$ decreases 
with increasing hole concentration as the magnetic correlations are weakened by 
the mobile holes. For $q_z=0$ (Fig.~\ref{sf2:ssf_wfAFSC}), it is observed 
that $S(\mathbf{q})$ increases with increasing hole concentration, the increase 
being maximum at $(\pi,\pi,0)$. This signals rapid suppression of interplanar 
AF long range correlations away from half-filling. 

\begin{figure}
\centering
\subfigure[]{\label{sf1:ssf_tz_Jz}
\includegraphics[width=1.6in]{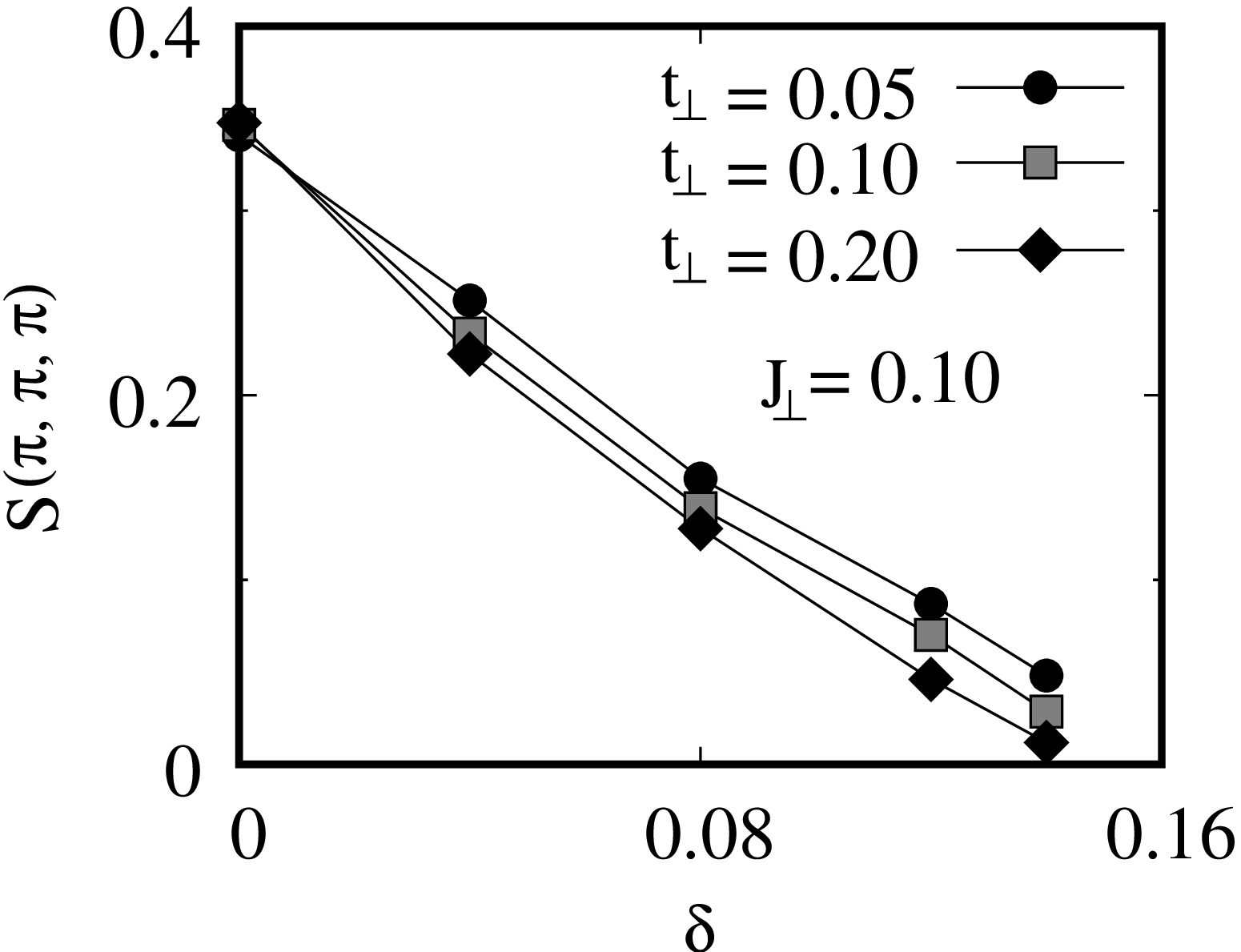}}  
\subfigure[]{\label{sf2:ssf_tz_Jz}
\includegraphics[width=1.6in]{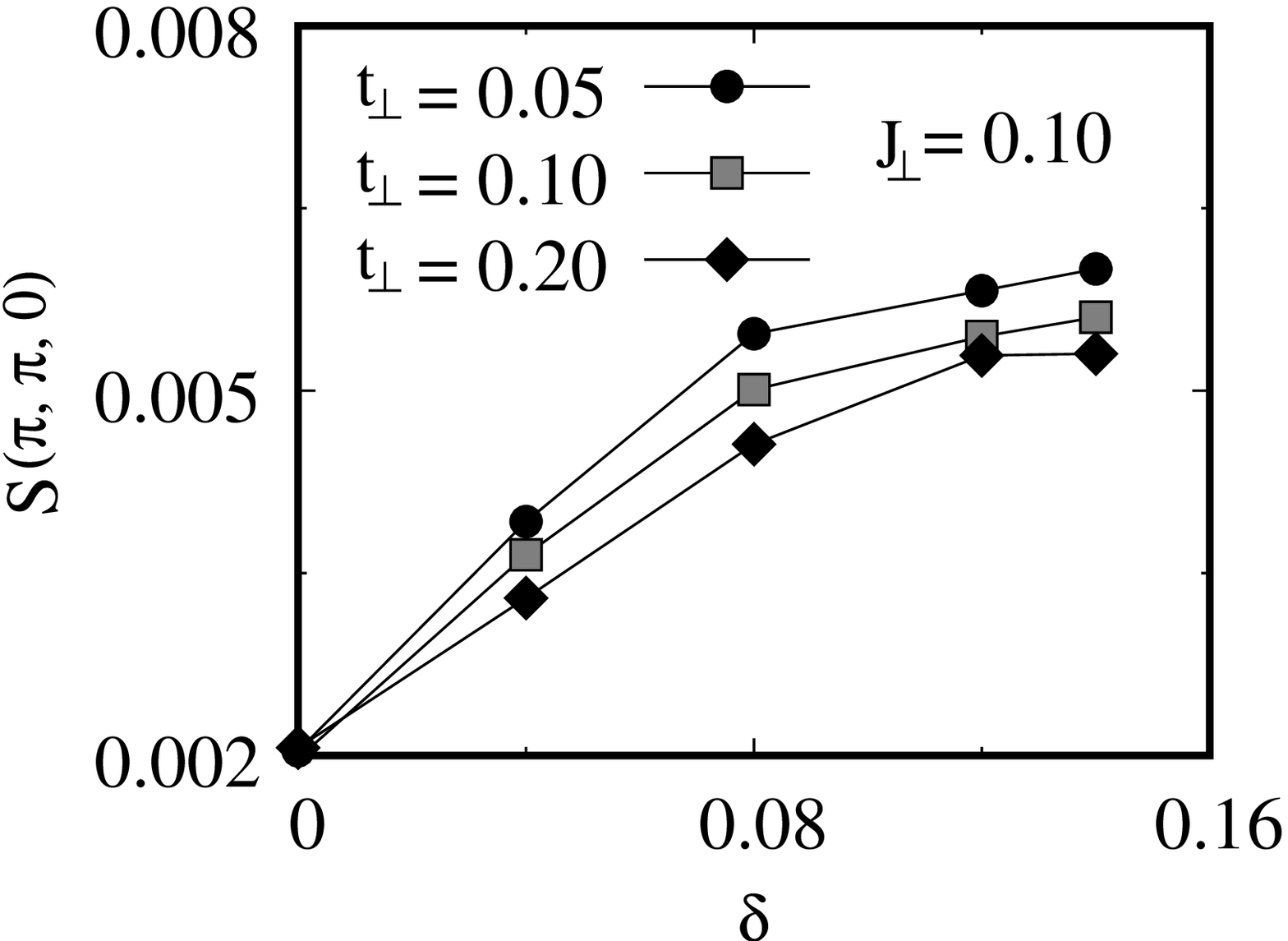}}  
\subfigure[]{\label{sf3:ssf_tz_Jz}
\includegraphics[width=1.6in]{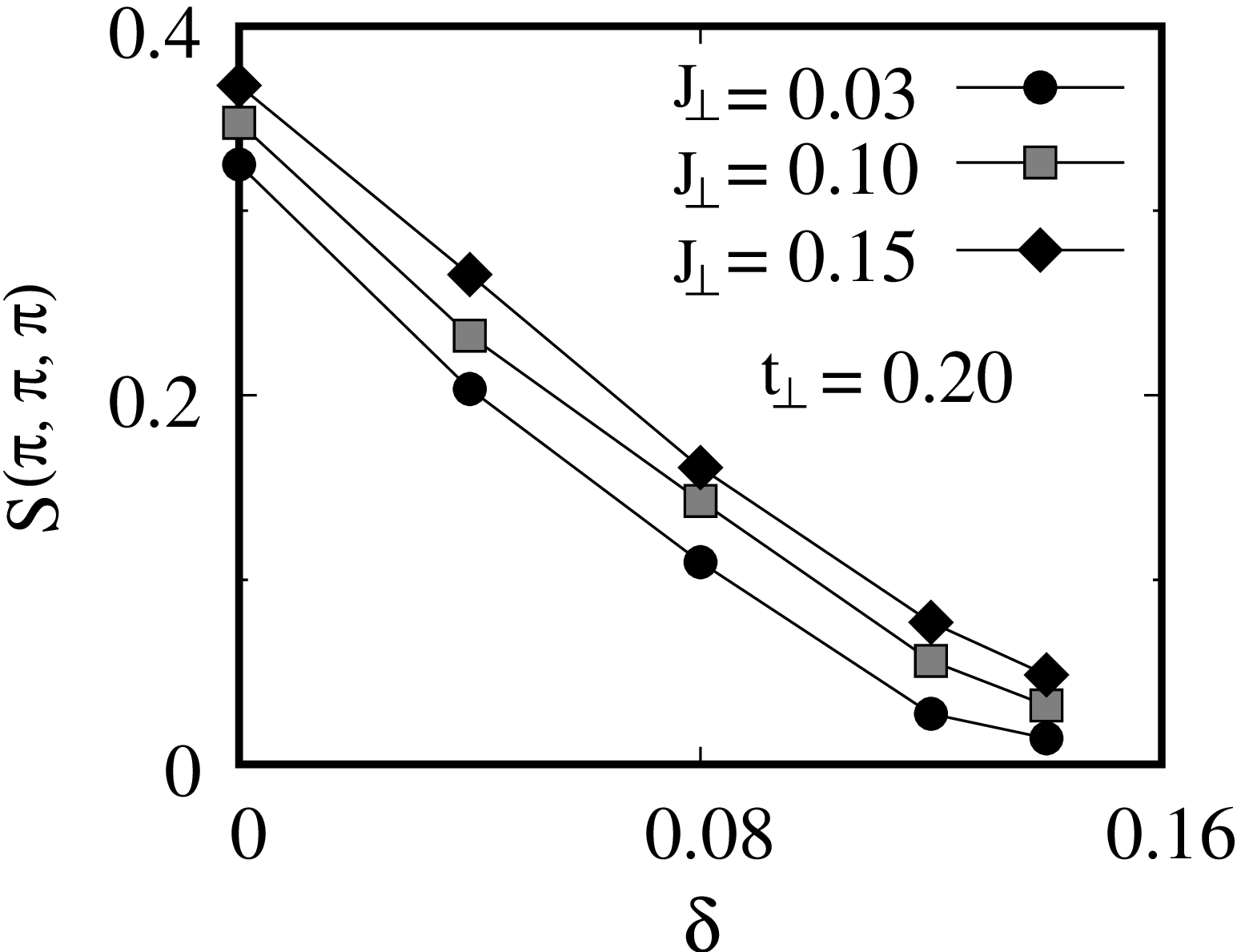}}  
\subfigure[]{\label{sf4:ssf_tz_Jz}
\includegraphics[width=1.6in]{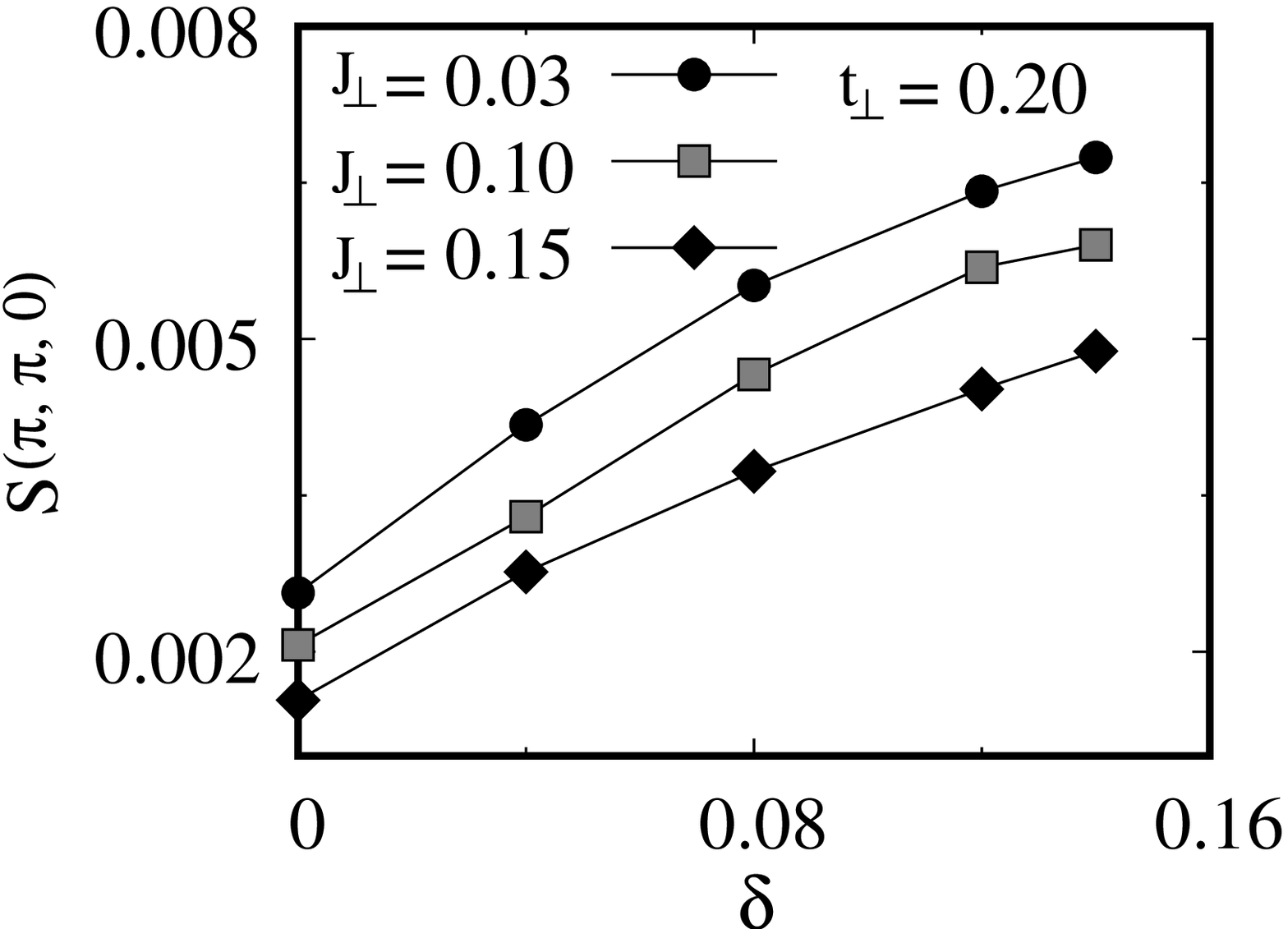}}  
\caption{$S(\pi,\pi,\pi)$ and $S(\pi,\pi,0)$ is plotted as a function 
of hole concentration for various values of interplanar parameters shown in 
figure, showing the effect of interplanar couplings on the AF long range order 
for the AF-SC state. (a), (b) involves variation of $t_{\perp}$ for a fixed
$J_{\perp}$ and (c), (d) contain variation of $J_{\perp}$ for a fixed 
$t_{\perp}$.}
\label{fig:ssf_tz_Jz}
\end{figure}

Next we incorporate the effect of interplanar couplings on the magnetic correlations.
Fig.~\ref{fig:ssf_tz_Jz} shows the variations of $S(\pi,\pi,\pi)$ and 
$S(\pi,\pi,0)$ with $t_{\perp}$ and $J_{\perp}$ both at and away from half-filling. 
In Figs.~\ref{sf1:ssf_tz_Jz} and (b), we let $t_{\perp}$ vary while keeping $J_{\perp}$
constant. It is seen that $t_{\perp}$ has no effect on magnetic correlations 
at half-filling, which is expected as hopping of electron  is forbidden at 
half-filling due to the no double occupancy constraint. Away from half-filling, 
$S(\pi,\pi,\pi)$ shows a decrease with increasing $t_{\perp}$ thus indicating 
AFLRO diminishes by larger interplanar hopping. Further, the decrease of 
$S(\pi,\pi,0)$ with increasing $t_{\perp}$ implies that the planar correlations 
are mainly affected. On the other hand, variation of interplanar exchange is 
found to have the reverse effect on the correlations. Fig~\ref{sf3:ssf_tz_Jz} 
shows that $S(\pi,\pi,\pi)$ increases with larger $J_{\perp}$ both 
at and away from half-filling. However $S(\pi,\pi,0)$ (Fig~\ref{sf4:ssf_tz_Jz})
decreases with increasing $J_{\perp}$. This is a reflection of the fact that
interplanar AF correlations are strengthened by larger $J_{\perp}$, while the
planar correlations are almost unaffected. This agrees with exact 
diagonalization results obtained for the bilayer \mbox{$t$-$J$} model 
\cite{EderPRB95}. Thus we conclude that larger $t_{\perp}$ reduces 
planar magnetic correlations away from half-filling, while larger $J_{\perp}$ 
enhances interplanar correlations both at and away from half-filling.

\subsection{Superconducting correlations}

Our next job constitutes of investigating SC correlations for 
both pure SC and coexisting AF and SC state. The SC correlation 
function is defined as,
\begin{equation}
F_{{\bm\alpha},{\bm\beta}}(\mathbf{r}-\mathbf{r}')=
\left\langle B^{\dag}_{\mathbf{r}{\bm\alpha}}
B_{\mathbf{r}'{\bm\beta}}\right\rangle
\end{equation}   
where $B$'s are pair operators and are represented by $B_{\mathbf{r}'{\bm\beta}}
=\frac{1}{2}(c_{\mathbf{r}'\uparrow}c_{{\mathbf{r}'+{\bm\beta}}\downarrow}-
c_{\mathbf{r}'\downarrow} c_{{\mathbf{r}'+{\bm\beta}}\uparrow})$ 
which annihilates a singlet pair on bond $(\mathbf{r}', \mathbf{r}'+{\bm\beta})$
and $B^{\dag}_{\mathbf{r}\alpha}$ creates one on $(\mathbf{r}, \mathbf{r}+
\alpha)$. $\bm\alpha$ and $\bm\beta$ are unit vectors connecting to nearest 
neighbours in the $x$, $y$ (planar) and $z$ (across the plane) directions. We 
have computed $F_{{\bm\alpha},{\bm\beta}}$ as a 
function of distance $|\mathbf{r}-\mathbf{r}'|$ for different hole 
concentrations corresponding to several choices for the interplanar parameters.
A very useful quantity in this connection is the SC order order parameter 
$\Phi$ which is obtained as\cite{ParamekantiPRB04} 
$F_{{\bm\alpha},{\bm\beta}}(\mathbf{r}-\mathbf{r}') \rightarrow\pm{\Phi}^2$ 
for large $|\mathbf{r}-\mathbf{r}'|$, with the sign being $+(-)$ for 
$\bm\alpha$ to be $\parallel(\perp)$ to $\bm\beta$ (both $\bm\alpha$ and 
$\bm\beta$ lie on a single layer). 

To examine interplanar SC correlations, we have calculated 
$F_{{\bm\alpha},{\bm\beta}}(\mathbf{r}-\mathbf{r}')$ taking both 
$\bm\alpha$, $\bm\beta$ to be along $z$-direction. The values obtained 
at all hole concentrations are very small (smaller than the error bars), 
{\it {i.e.}} negligible in comparison to the planar correlation values.
The result appears as no surprise as the pure $d$-wave state which is found to 
most suitably describe pairing symmetry for a bilayer, contains no
significant interplanar SC correlations. 

\begin{figure}[htb]
\centering
\includegraphics[width=3.0in]{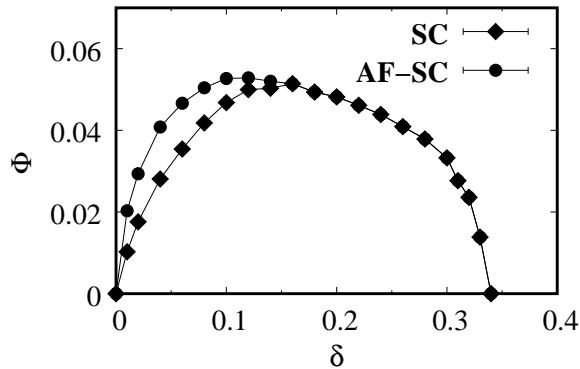}
\caption{SC order parameter, $\Phi$ shown as a function of hole concentration, 
$\delta$ for the AF-SC and pure SC state. Here the interplanar parameters are 
chosen as $t_\perp = 0.20$ and $J_{\perp}=0.10$.}
\label{fig:scorder_wfA_tz0.20_Jz0.10}
\end{figure}

Next we show that the planar SC correlations are stronger in the AF-SC state 
in the region of hole doping ({\it {i.e.}} underdoped) in which the two phases 
coexist in comparison to that in the pure SC state. 
Fig.~\ref{fig:scorder_wfA_tz0.20_Jz0.10} shows the SC order parameter, $\Phi$ 
corresponding to planar $F_{{\bm\alpha},{\bm\beta}}(\mathbf{r}-\mathbf{r}')$ 
({\it {i.e.}} $({\bm\alpha},{\bm\beta})$ along ($x$,$y$) as function of 
$\delta$ for both the pure SC and the AF-SC state. 
The figure clearly shows that SC correlations are stronger in the coexisting
phase than in the pure SC state. It is very interesting to note that the 
coexisting phase which is energetically favourable in the underdoped region, gives 
rise to not only {\em {stronger AF correlations}} but also {\em {enhanced SC 
correlations}} as compared to that in pure SC state. This is another important 
result of our paper.

\begin{figure}[htb]
\centering
\includegraphics[width=3.0in]{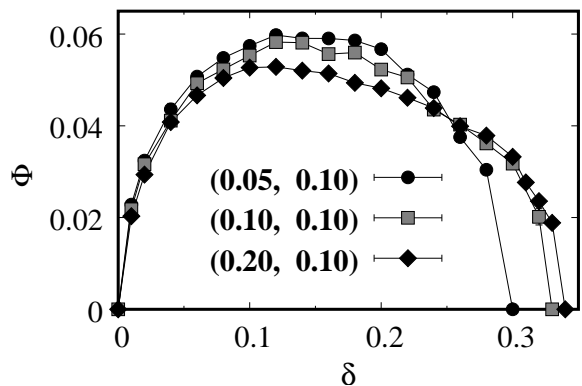}
\caption{SC order parameter, $\Phi$ versus hole concentration, $\delta$ 
for different values of $t_\perp$  with $J_{\perp} = 0.1$ 
The numbers in parentheses shown in figure are the values of 
($t_\perp$, $J_{\perp}$).}
\label{fig:scorder_wfAFSC_tzA_JzA}
\end{figure}

Next to estimate the effects of interplanar couplings on the SC correlations in
a plane, we calculate the SC order parameter, $\Phi$ as a function of hole 
concentrations for different choices of  $t_{\perp}$ and $J_{\perp}$. 
The results are shown in Fig.~\ref{fig:scorder_wfAFSC_tzA_JzA} with $t_{\perp}$
varying from $0.05$ to $0.2$ and $J_{\perp} = 0.1$. 
From the figure one observes that the SC correlations varies 
significantly and differently in the optimally doped and the overdoped
regimes with the interplanar hopping $t_\perp$.
In the underdoped region, the effect of $t_{\perp}$ is small. At optimal values of
doping, {\it {i.e.}} in the peak region of order parameter $\Phi$, larger 
interplanar hopping reduces planar 
SC correlations strongly. However,  just the opposite behaviour is 
observed in the overdoped region. As shown in 
Fig.~\ref{fig:scorder_wfAFSC_tzA_JzA}, for $t_{\perp} = 0.05$, the 
optimally doped region is marked by largest $\Phi$, indicating greater stability
of the SC state for smaller $t_{\perp}$. Whereas the critical hole 
concentration, $\delta_c$ at which the phase transition from SC to normal 
state takes place, is $\sim 0.30$ for $t_{\perp} = 0.05$, while for $t_\perp 
= 0.10$ and $0.20$, $\delta_c$'s are obtained as $0.33$ and $0.34$ respectively.
Thus increasing the interlayer hopping in a bilayer 
results in greater stability of the SC state in the overdoped region. 
With regard to the impact due to interplanar exchange, $J_{\perp}$, our 
calculations show that $J_\perp$ has no perceptible effect (figures not shown 
here) on the SC correlations at least for the values of $J_\perp$ considered 
here. 

\begin{figure}[htb]
\centering
\includegraphics[width=3.0in]{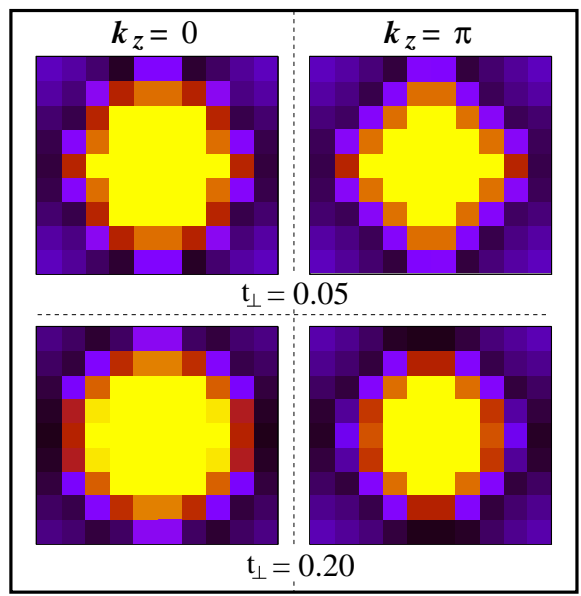}
\caption{(color online) Plots of $n(\mathbf {k})$ projected on a two-dimensional
($k_x,\,k_y$) plane where  white (yellow) and gray (magenta) colors 
represent the highest and lowest spectral weights respectively. The left panel 
shows $n(k)$ for $k_z=0$ and the right panel corresponds to $k_z=\pi$, for 
$t_{\perp} = 0.05$ (top) and $0.20$ (bottom). The hole concentration here is
$\delta=0.32$.}
\label{fig:fermisurface}
\end{figure}

The asymmetric behaviour shown by superconducting correlations in the
optimally doped and the overdoped regime as $t_{\perp}$ is increased 
from $0.05$ to $0.20$ constitutes another key finding of this paper.
While the reduction of planar SC correlations with increasing 
$t_{\perp}$ as observed in optimally doped region is expected due to the 
fact that larger hopping have a disrupting on the pairing, the reason
for the opposite behaviour in the overdoped regime  where SC correlations
are enhanced by larger $t_{\perp}$, is not immediately obvious. To explain 
this we look at momentum distribution for two 
different values of $t_{\perp}$, {\it viz.} $0.05$ and $0.20$. 
Fig.~\ref{fig:fermisurface} shows the
2D projection of momentum distribution function, $n({\mathbf k})$ with
$n({\mathbf k})=\langle c_{\mathbf k\sigma}^{\dag}c_{\mathbf k\sigma}\rangle$
for $\delta=0.32$. At this hole doping value there is a large variation in
SC order parameter for the two values of $t_{\perp}$ (actually SC order is
vanishing small at $\delta = 0.32$ for $t_{\perp} = 0.05$)
It is seen from the plots
that there a significant variation of $n(\mathbf k)$ weights in the two
momentum planes, {\it viz.} the $k_z=0$ and $k_z=\pi$, as $t_{\perp}$ is
varied. The spectral weight shifts from the $k_z=\pi$ plane to the $k_z=0$ plane
as $t_{\perp}$ is increased from $0.05$ to $0.20$. Thus for larger $t_{\perp}$,
occupation of the pairs with $k_z=0$ is higher in
the overdoped region. This favours the stability of SC phase as the
$k_z=0$ pairs are planar and contribute in development of the SC order.
The $k_z=\pi$ pairs are interplanar and hence are not expected to be
key players as the SC correlations are essentially planar. We believe that this
transfer of weight from $k_z=\pi$ to $k_z=0$ for larger $t_{\perp}$ helps in
stabilizing the SC order in the overdoped region even though the disrupting
effect due to larger interplanar hopping persists. In the optimally doped 
region, the momentum distribution profiles for different values of $t_{\perp}$ 
are found to be almost identical (plots not shown here) and hence the only 
effect of $t_{\perp}$ is to reduce the SC correlations.

Regarding the finite size effects in our results, we would like to mention here that
the magnitudes of various quantities calculated do show some dependence on the size
of the lattice. The dependence of energy on lattice size is elucidated in details
in Ref.~\onlinecite{MedhiPHC06}.
However the main features of the key results here, e.g. the coexistence of AF and 
SC order,  the effect of interlayer parameters on the properties etc.\ will remain 
qualitatively same with lattice size.

\section{Conclusion}
We summarise our main results obtained using variational calculations for a 
\mbox{$t$-$J$} bilayer as follows - coexistence of AF-SC is 
found to be more stable than the
pure ($d$-wave) SC state at low values of doping ($\delta < 0.14$). Beyond 
this, of course, the SC state is found to have the lowest energy and remains
stable upto a hole concentration that is more than that obtained for the 
two-dimensional square lattice. Further,
a detailed analysis of magnetic and superconducting properties 
yields  the coexisting phase, not only energetically stable, but also supports
a stronger AF and SC correlations. The third and possibly the most important 
result emerges when the effects of interplanar coupling parameters are
invoked for discussion and it appears in the form of an asymmetry in 
the optimally and the overdoped region where the planar SC correlations 
are found to be more stable for a smaller and larger interplanar hopping 
respectively. However the interplanar exchange does not play a decisive role
in SC correlations.

Some of the other results obtained by us include  a comparison of the 
nature of magnetic correlations between the pure SC state and the AF-SC state. 
It is found that strong planar and interplanar AF correlations exists in 
the AF-SC state at and slightly away from half-filling, whereas in the pure 
SC state the magnetic correlations are very weak. We have also discussed the 
effects of interplanar coupling on magnetic correlations in the 
AF-SC state where larger $t_{\perp}$ reduces planar magnetic correlations 
whereas larger $J_{\perp}$ enhances interplanar magnetic correlations.

\begin{acknowledgments}
One of us (AM) would like to thank CSIR, India for financial support under the
Grant No - F.NO.9/731(022)/2003-EMR-.
\end{acknowledgments}


\begin{thebibliography}{unsrt}
\bibitem{Weakcorrelations} $S(\pi,\pi,\pi)$ would have significantly differed 
from $S(\pi,\pi,0)$ had the interplanar magnetic correlations been strong.
\bibitem{CompareSSF} In Ref.~\onlinecite{EderPRB95}, $S(\mathbf{q})$ is 
obtained by considering $\langle \mathbf{S}_i.\mathbf{S}_j\rangle$ instead of
$\langle S_i^z.S_j^z\rangle$ in Eq.~\ref{eq:ssf}. Hence $S(\mathbf{q})$,
according to our definition, has to be scaled by $\frac{1}{3}$ owing to the 
spin rotational invariance.
\end{thebibliography}
\end{document}